\documentclass[twoside]{article}
\usepackage{fleqn,espcrc2}

\usepackage{graphicx}
\usepackage{epsfig}
\usepackage[figuresright]{rotating}

\newcommand{\AmS}{{\protect\the\textfont2
  A\kern-.1667em\lower.5ex\hbox{M}\kern-.125emS}}

\def\taub{{\overline{\tau}}}
\def\etab{{\overline{\eta}}}
\def\bk{{\bf{k}}}
\def\dst{\displaystyle\strut}
\def\be{\begin{equation}}
\def\ee{\end{equation}}
\def\bea{\begin{eqnarray}}
\def\eea{\end{eqnarray}}
\def\l({\left(}
\def\r){\right)}

\def\I#1{\int d^3#1}
\def\bx{{\bf{x}}}
\def\bp{{\bf{p}}}
\def\bk{{\bf{k}}}
\def\bpi{{\bf{\pi}}}
\def\bxi{{\bf{\xi}}}
\def\bq{{\bf{q}}}
\def\br{{\bf{r}}}
\def\axd{\hat{ a}^{\dag} (\bx)}
\def\apd{\hat a^{\dag} (\bp)}
\def\ax{\hat{ a}^{} (\bx)}
\def\ap{\hat  a^{} (\bp)}

\def\ri{\right)}
\def\lef{\left(}
\def\om{\omega}

\def\dst{\displaystyle{\phantom{|}}}
\def\ov{\over\dst}

\title{Particle interferometry, \\
	binary sources and oscillations in two-particle correlations}
\author{T. Cs\"org\H{o}\\
MTA KFKI RMKI\\
	   H - 1525 Budapest 114, POB 49\\
	   Hungary}
\begin{document}
\begin{abstract}
The basics and the  formalism of Bose-Einstein correlations
is briefly reviewed. The invariant Buda-Lund form is
summarized.  Tools are presented that can be utilized
in a model-independent search for non-Gaussian structuctures in the
two-particle Bose-Einstein correlation functions.
The binary source formalism of particle interferometry is
presented and related  to oscillations in the two-particle
Bose-Einstein and Fermi-Dirac correlation functions. 
The frequency of the observed oscillations 
in the NA49 two-proton correlation function
in Pb+Pb collisions at CERN SPS energies
is explained with the help of the reconstructed space-time
picture of particle production and the binary nature of the proton
source in this reaction. 
\vspace{1pc}
\end{abstract}
\maketitle
%\bigskip
%\rightline{\large\it 
%`` Never test for an error condition }
%\rightline{\large\it 
%you don't know how to
%handle.  "}
%\rightline{\large\sc Steinbach's guideline }
%\rightline{\large\sc
%for systems programming }
%\bigskip
%\bigskip
%----------------------------------------------------------
% 16 x 4 = 64
%

%-----------------------------------------------------------
%	1. Intro
\section{Introduction}
	Essentially, intensity correlations appear due to the 
	Bose-Einstein or Fermi-Dirac symmetrization of the 
	two-particle final states of identical bosons or fermions,
	in short, due to quantum statistics.
	Intensity correlations were discovered for the first time
	in radio astronomy by R. Hanbury Brown and R. Q. Twiss,
	~\cite{HBT,HBT56}	and were utilized to determine
	the angular diameter of main sequence stars,
	the HBT effect.
	In particle physics, the enhancement of
	the production of identical pions at small angular
	separations was discovered by Goldhaber, Goldhaber, Lee and Pais
	(GGLP) in refs.~\cite{gglp,gglp0}.

	In this contribution, some comments are made on the frequently
	invoked fully thermal and fully chaotic limiting cases
	and  it is shown why the Andersson-Hofmann model
	corresponds to neither of these cases. 
	Certain subtle aspects of two-particle
	Bose-Einstein and Fermi-Dirac correlations are highlighted,
	that can be utilized in experimental searches for new,
	non-Gaussian structures in the two-particle 
	quantum statistical correlation functions.
	Some of the material discussed here is described in greater
	details in the recent review paper~\cite{cs-nato}. Various
	complementary aspects of the field were reviewed recently 
	in refs.
	~\cite{bengt,zajc,jacak-heinz,uli_urs,weiner_rep,kittel,muller-qm99}.

\subsection{Basics of intensity correlations}
	
	The simplest derivation of the HBT/GGLP/Bose-Einstein
	correlation effect is as follows:	
	suppose that a particle pair is observed, one with momentum
	$k_1$ the other with momentum $k_2$. The 
	amplitude of pair emission 
	has to be symmetrized over the unobservable 
	variables, in particular over the points of emissions
	$x_1$ and $x_2$.
	If Coulomb, strong or other final state
	interactions can be neglected, 
	the amplitude of such a final state is proportional to
\be
	A_{12} \propto \frac{1}{\sqrt{2}}  
		\,\,	[\, {\rm e}^{i k_1 x_1 + i k_2 x_2} \pm
			{\rm e}^{i k_1 x_2 + i k_2 x_1} \,],
\ee
	where $+$ sign stands for bosons, $-$ for fermions.
	If the particles are emitted in an incoherent manner,
	from a non-expanding source, the observable 
	two-particle spectrum is proportional
	to
\be
	N_2(k_1,k_2) \propto 
		\int dx_1 \rho(x_1) 
		\int dx_2 \rho(x_2) \,\, |A_{12}|^2
\ee
	and the resulting two-particle intensity correlation function is
\be
	C_2(k_1,k_2)  =   \frac{N_2(k_1,k_2)}{N_1(k_1) N_2(k_2)} \,
	= 1 \pm |\tilde \rho(q) |^2, \label{e:rhoq}
\ee
	that carries information about the Fourier-transformed space-time 
	distribution of the particle emission 
\be
	\tilde\rho(q) = \int dx \,\, {\rm e}^{i q x} \,\, \rho(x).
\ee 
	as a function of the relative four-momentum $q \equiv q_{12} = k_1 - k_2$.
	
	Although this derivation is over-simplified, the above result
	can be translated to the  more generally valid identity:

\be
	\langle a_1^{\dagger}	a_2^{\dagger}	a_2	a_1 \rangle \, = \,
	\langle a_1^{\dagger}	a_1 \rangle \, 
	\langle a_2^{\dagger}	a_2 \rangle  \pm
	\langle a_1^{\dagger}	a_2 \rangle \, 
	\langle a_2^{\dagger}	a_1 \rangle 
	, \label{e:chaos}
\ee
	where $a_i^{\dagger} $ and $a_i$ are the creation and annihilation
	operators for identical particles with 4-momentum $k_i$, $(i = 1,2)$ and
	$\langle O \rangle = Tr \rho O $ stands for the expectation value
	of the operator $O$ in a system characterized by a density matrix
	$\rho$.

	A more sophisticated derivation of this and the corresponding
	$n$-particle Bose-Einstein correlation functions is reviewed
	e.g. in ref.~\cite{cs-nato}. Here we highlight  only
	the essential properties of various statistical features
	of the quantum statistical correlations, hence the mathematical
	complications will be limited to the minimal level.
	
	As compared to the idealized case when quantum-statistical correlations
	are negligible (or neglected), Bose-Einstein or Fermi-Dirac
	correlations modify the momentum distribution
	of the hadron pairs in the final state by a weight factor 

\be
	\langle a_1^{\dagger}	a_2^{\dagger}	a_2	a_1 \rangle =
%& = & 
	\langle a_1^{\dagger}	a_1 \rangle \, 
	\langle a_2^{\dagger}	a_2 \rangle 
	% \times \nonumber \\
% && \hspace{-1truecm}	
\{ 1 \pm  |\langle e^{ i q_{12} x } \rangle |^2 \}.
	\nonumber 
% \\
%	&& 
	 \label{e:chaosweight}
\ee

	Note that eq.~(\ref{e:chaos}) is valid for chaotic 
	(for example, locally thermalized) systems 
	and it is in a very good agreement with
	the detailed experimental and theoretical investigations of 
	quantum statistical correlation functions in high energy 
	heavy ion collisions.

	However, eq.~(\ref{e:chaos}) is not valid in case of a coherent
	particle emission. In case of a brehmstrahlung - like or laser -
	like coherent radiation, eq.~(\ref{e:chaos}) is replaced by
\be
	\langle a_1^{\dagger}	a_2^{\dagger}	a_2	a_1 \rangle \, = \,
	\langle a_1^{\dagger}	a_1 \rangle \, 
	\langle a_2^{\dagger}	a_2 \rangle . 
	\label{e:coh}
\ee

	Thus the non-trivial quantum statistical intensity correlations
	appear in chaotic, thermal like sources. They correspond to 
	the second term of eq.~(\ref{e:chaos}). Second order optical
	coherence is defined by the vanishing value of these
	kind of exchange terms, compare eq.~(\ref{e:coh}) 
	with eq.~(\ref{e:chaos}).
 
	Thus the applicability of the starting point of many detailed
	derivations, eq.~(\ref{e:chaos} ) is limited to chaotic, 
	thermal-like sources. 
	If some degree of coherence 
	is preserved during particle production,
	eqs.~(\ref{e:chaos},\ref{e:chaosweight}) become invalid. 
	This is the case of not only in case of a coherent, 
	laser-like radiation but also in case of 
	the Andersson-Hofmann model of Bose-Einstein
	correlations in $e^+ e^-$ collisions.

	Already at this basic level, one can realize the uniqueness of
	the Andersson-Hofmann model of a two-particle
	Bose-Einstein correlations in $e^+ e^-$ annihilation.

\section{The Andersson-Hofmann model }
The hadronic production in $e^+ e^-$ annihilations is usually
considered to be a basically coherent process and therefore no
Bose-Einstein effect was expected in these reactions 20 years ago. 
It was also thought that the
hadronic reactions should be of a more chaotic nature giving rise to a sizable
effect. It was even argued that the strong ordering in rapidity,
preventing neighbouring $\pi^-\pi^-$ or $\pi^+ \pi^+$ pairs,
would drastically reduce the effect~\cite{veneciano}. Therefore
it was a surprise when G. Goldhaber at the Lisbon Conference in 1981
\cite{goll} presented data which showed that correlations between
identical particles in $e^+ e^-$ annihilations were very similar
in size and shape to those seen in hadronic reactions, see the review paper 
ref.~\cite{bengt} for further details.

\begin{figure}
\begin{center}
\epsfig{file=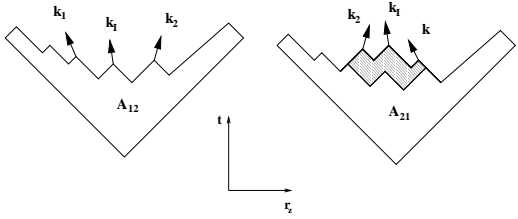,width=1.1in,angle=270}
\end{center}
\caption{\label{f:andersson} 
Andersson - Hofmann interpretation of 
Bose-Einstein correlations in the Lund string model.
$A_{12}$ denotes the space-time area of a colour field enclosed
by the quark loop in $e^+ e^-$ annihilation. 
Two particles 1 and 2 are separated by the
intermediate system $I$. When the particles 1 and 2 are identical,
the configuration in the left side is indistinguishable from
that of the right side, and their amplitudes for production must
be added. The probability of production will depend on the
difference in area $\Delta A = A_{12} -A_{21}$, shown as the hatched area.
}
\end{figure}

The Bose-Einstein correlation effect,
{\it a priori} unexpected for a coherent process, has been given
an explanation within the Lund string model by 
B. Andersson and W. Hofmann~\cite{andersson}. The space-time structure 
of an $e^+ e^-$ annihilation is shown
 for the Lund string model \cite{lundm} in
Figure~\ref{f:andersson}.
The probability for a particular final state is given by
the Lund area law 
\begin{equation}
{\rm Prob} \sim {\rm phase space} \cdot \exp(-bA),\label{e:pb24}
\end{equation}
where $A$ is the space-time area spanned by the string before it
breaks and $b$ is a parameter. The classical
string action is given by $S= \kappa A$, where $\kappa $ is the
string tension. It is natural to interpret the result in
eq.~(\ref{e:pb24}) as resulting from an imaginary part of the
action such that
\bea
S &= & \xi A,\\
{\it Re} \xi &= &\kappa,\\
{\it Im} \xi &= &b/2,
\eea
and an amplitude $M$ given by
\begin{equation}
M \sim \exp(iS),
\end{equation}
which implies
\begin{equation}
{\rm Prob} \sim \mid M \mid^2 \sim \exp(-bA).
\end{equation}

Final states with two identical particles are indistinguishable
and can be obtained in different ways. Suppose that the two
particles indicated as 1 and 2 on Fig.~\ref{f:andersson} are identical,
then 
the
hadron state in the left panel can be considered as being the same as 
that in the right panel  (where 1 and 2 are interchanged).
The amplitude should, for bosons, be the sum of two terms
\be
M \sim M_{12} + M_{21}  \,  = \,
\exp[i\xi A_{12}] +\exp[i \xi A_{21}]
\ee
where $A_{12}$ and $A_{21}$ are the two string areas, giving a
probability proportional to
\bea
\mid M \mid^2 & \sim & [\exp(-bA_{12}) + \exp(-bA_{21})]\times \nonumber \\
&& [1 +{\cos(\kappa\Delta A)\over \cosh(b\Delta A/2)}]
\eea
with $\Delta A \equiv A_{12} - A_{21}$.  The magnitudes of $\kappa $
and $b$ are known from phenomenological studies. 
The energy per unit length of the string is given by $\kappa
\approx 1$ GeV/fm, and $b$ describes  the breaking of the string
at a constant rate per unit area, $b/\kappa^2 \approx 0.7 $ GeV$^{-2}$
\cite{lundm}. The difference in space-time area $\Delta A$ is
marked as the hatched area in Fig.~\ref{f:andersson}. It can be
expressed by the $(t,r_z)$ components $(E,k)$ 
of the  four-momenta of the two
identical particles 1 and 2, and the intermediate system $I$:
$$
\Delta A =  [E_2 k_1 - E_1 k_2 + E_I(k_1 - k_2) - k_I(E_1-E_2)]/\kappa^2\nonumber\nonumber
$$
To take into account also the component transverse to the string
a small additional term is needed. The change in area $\Delta A$
is Lorentz invariant to boosts along the string direction and is
furthermore approximately proportional to 
$Q = \sqrt{ - (k_1 - k_2)^2}$.

The interference pattern  between the amplitudes will be
dominated by the phase change of $\Delta \Phi = \kappa \Delta A$.
It leads to a Bose-Einstein correlation which, as a function of
the four-momentum transfer, reproduces the data well but shows a
steeper dependence at small $Q$ than a Gaussian function. A
comparison to TPC data confirmed the existence of such a 
steeper than Gaussian dependence  on $Q$,
although the statistics at the small $Q$-values 
did not allow a firm conclusion~\cite{bengt,TPC}.

An essential feature of the Andersson - Hofmann model is that

\bea
	\langle a_1^{\dagger}	a_2^{\dagger}	a_2	a_1 \rangle & = & 
	\langle |M_{12} + M_{21}|^2 \rangle \, \nonumber \\ 
	& \ne &  \langle M_{11} \rangle \langle M_{22} \rangle + 
	| \langle M_{12} \rangle|^2
\eea
and
\bea
	\langle a_1^{\dagger}	a_2^{\dagger}	a_2	a_1 \rangle & = & 
	\langle |M_{12} + M_{21}|^2 \rangle \nonumber \\
  & \ne  & \langle M_{11}  \rangle \langle M_{22} \rangle
\eea
i.e. the source is neither chaotic, nor fully coherent. 
In case of a well defined intermediate system I, the phase difference
between the two amplitudes is also well defined; however, I is randomly
varying from event to event, that leads to a variation of the phases
and the onset of a chaotic like behaviour. As the phase difference 
 $ \propto \Delta A$ is, however, not a uniformly distributed random variable
in the Lund model, a residual phase correlation survives in the two-particle
	Bose-Einstein correlation function, that cannot be obtained otherwise,
neither in fully chaotic , nor in fully coherent systems.

In the subsequent parts, we consider only fully chaotic systems,
relevant for the study of hadron-hadron, hadron-nucleus and nucleus-nucleus
interactions at high energies. 

%-----------------------------------------------------------
%	7. Large Q expansion (Buda-Lund)
\section{Buda-Lund particle interferometry }
\label{s:bl}

\def\vp{{\bf k}}
\def\vq{{\bf q}}
\def\vk{{\bf k}}
\def\vK{{\bf K}}
\def\vx{{\bf x}}
\def\vy{{\bf y}}
\def\uk{{|{\bf k}|}}
\def\bc{\begin{center}}
\def\ec{\end{center}}
\def\De{\Delta\eta}
\def\Det{\Delta\eta_T}
\def\Des{\Delta\eta_*}
\def\Dk{\Delta k}
\def\Dt{\Delta\tau}
\def\Dy{\Delta y}
\def\t0{\tau_0}
\def\tl{\tau_L}
\def\ch{\cosh}
\def\sh{\sinh}
\def\bx{{\bf{x}}}
\def\bp{{\bf{p}}}
\def\bk{{\bf{k}}}
\def\bK{{\bf{K}}}
\def\bK{{\bf{K}}}
\def\bQ{{\bf{Q}}}
\def\bpi{{\bf{\pi}}}
\def\bxi{{\bf{\xi}}}
\def\bq{{\bf{q}}}
\def\br{{\bf{r}}}
\def\bak{{\bf K}}
\def\dek{{\bf \Delta k}}
\def\xb{{\overline{x}}}
\def\tb{{\overline{t}}}
\def\rb{{\overline{r}}}
\def\nb{{\overline{n}}}
\def\etab{{\overline{\eta}}}
\def\taub{{\overline{\tau}}}

%
% Now come the definitions for the new notation scheme
%
\def\D{\Delta}
\def\o{{out}}
\def\s{{side}}
\def\e{{\eta}}
\def\bdk{{\bf \Delta k}}
\def\rl{R_l^2}
\def\ro{R_{o}^2}
\def\rs{R_{s}^2}
\def\rol{R_{ol}^2}
\def\rpa{R_{\parallel}^2 }
\def\rpe{R_{\perp}^2 }
\def\rta{R_{\tau}^2 }
\def\BL{{Buda}{-}{Lund}~} 

	The $n$-particle Bose-Einstein correlation function of
	is defined as the 
	ratio of the $n$-particle invariant momentum 
	momentum distribution divided by an $n$-fold product
	of the single-particle invariant momentum distributions.
	Hence these correlation functions are boost-invariant. 

	The invariant Buda-Lund parameterization (or BL in short)
	deals with a boost-invariant, multi-dimensional
	characterization of the building blocks
	$\langle a_{{\bf k}_1}^{\dagger} a_{{\bf k}_2}
	\rangle $ of arbitrary high order   Bose-Einstein correlation
	functions.
	The BL parameterization was developed by the Budapest-Lund
	collaboration in refs.~\cite{3d,mpd95}.

\begin{figure}
\begin{center}
\vspace{1cm}
\epsfig{file=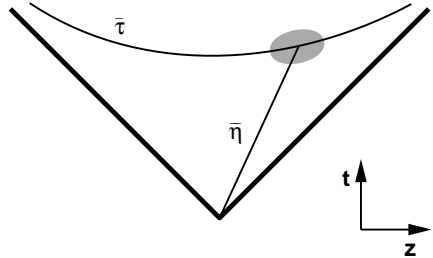,width=2.2in,angle=0}\\
\end{center}   
\caption{
	Space-time picture of particle emission for
        a given fixed mean momentum of the pair.
        The mean value of the proper-time
        and the  space-time rapidity    distributions
        is denoted by $\taub$ and  $\etab$.
        As the rapidity of the produced particles changes from the
        target rapidity to the projectile rapidity
        the $[\taub(y),\etab(y)]$ variables scan the
        surface of mean particle production in the $(t,r_z)$   plane.
}
\label{f:bl}
\vspace{0.3cm}
\end{figure}

	The essential part of the BL 
	is an invariant decomposition of the 
	relative four-momentum $q$ in the 
	$\langle \exp(i q x) \rangle$ 
	factor of eq.~(\ref{e:chaosweight}) into a temporal, 
	a  longitudinal and two transverse 
	relative momentum components.   This decomposition
	is obtained with the help of a time-like vector
	in the coordinate space, that characterizes the center
	of particle emission in space-time, see Fig.~\ref{f:bl}.

	Although the BL parameterization was introduced in ref.~\cite{3d} for
	high energy heavy ion reactions, it can be used for
	other physical situations as well, where
	a dominant direction of an approximate boost-invariant
	expansion of the particle emitting source can be identified 
	and taken as the longitudinal direction $r_z$. For example,
	such a direction is the thrust axis of single jets 
	or of back-to-back two-jet events
	in case of high energy particle physics. 
	For longitudinally almost boost-invariant systems, 
	it is advantageous to introduce the boost invariant
	variable $\tau$ and the space-time rapidity $\eta$,
\bea
	\tau &	= & \sqrt{t^2 - r_z^2}, \\
	\eta & = & 0.5 \log\left[(t+r_z)/(t-r_z)\right] .
\eea
	Similarly, in momentum space one introduces the transverse mass
	$m_t$ and the rapidity $y$ as
\bea
	m_t & = & \sqrt{E^2 - p_z^2}, \\
	y & = &  0.5 \log\left[(E+p_z)/(E-p_z)\right] .
\eea
	The source of particles is characterized in 
	the boost invariant variables $\tau$, $m_t$ and $\eta - y$. 
	For systems that are only approximately boost-invariant, 
	the emission function may also depend on the  
	deviation from mid-rapidity, $y_0$.
	The scale on which the approximate boost-invariance breaks down
	is denoted by $\Delta \eta$, a parameter that is related to the 
	width of the rapidity distribution.

	The correlation function is defined with the help of the
	Wigner-function formalism, the intercept 
	parameter $\lambda_*$ is introduced in the core-halo picture
	see refs.~\cite{3d,chalo,cs-nato} for further details.
	In the following, we evaluate the Fourier-transformed
	Wigner functions, that provide the building block
	for arbitrary high order Bose-Einstein 
	correlation functions. 
	We  assume for simplicity that the core is fully incoherent. 

	Such a pattern of particle production is visualized
	in Fig. ~\ref{f:bl}.
	
	If the 
	production of particles with a  given  rapidity $y$ 
	is limited to a narrow region in 
	space-time around $\etab$ and $\taub$, and 
	the sizes of the effective source are sufficiently
	small (if the Bose-Einstein
	correlation function is sufficiently broad),
	the exponent of the  $\exp( i q \cdot \Delta x)$ factor 
	of the Fourier-transformation
	can be decomposed in the shaded region in Fig.~\ref{f:bl} 
	as
\bea
	q_0 t -  q_z r_z
	& \simeq  &
	Q_= ( \tau - \overline{\tau} )
		- Q_{\parallel} \taub (\eta -\overline{\eta}),
	\label{e:blexp} \\
	q_x r_x +  q_y r_y
	 & \equiv & 
	Q_: r_:  +  Q_{..}  r_{..} .
		\label{e:bltexp}
\eea
	The invariant {\it temporal}, {\it parallel},
	{\it side}ward, {\it out}ward (and
	{\it perp}endicular )
	relative momentum components are defined, respectively, as  
\bea
	Q_= & = &   
		q_0 \cosh[\etab] - q_z \sinh[\etab],
		 \label{e:q=} \\
	Q_{\parallel} & = & 
		q_z \cosh[\etab] - q_0 \sinh[\etab],
		\label{e:qpar} \\
	Q_{..} & = &  (q_x K_y - q_y K_x)/\sqrt{K_x^2 + K_y^2}, \\ 
	Q_{:} & = & (q_x K_x + q_y K_y)/\sqrt{K_x^2 + K_y^2}, \\ 
	Q_{\perp} & = & \sqrt{ q_{x}^2 + q_{y}^2} = \sqrt{Q_{:}^2+Q_{..}^2}.  	
\eea
	The timelike  normal-vector
	 	$\nb$ 
	indicates an invariant direction of
	the source in coordinate space~\cite{3d}. 
	It is parameterized as
	$\nb^{\mu} = (\cosh[\etab],0,0,\sinh[\etab])$, 
	where $\etab$ is a mean space-time 
	rapidity ~\cite{3d,3d-cf98,mpd95}. 
	The parameter $\etab$ is one of the fitted parameters in
	the BL type of decomposition of the relative momenta.	
	The above equations are invariant, they  can
	be evaluated in any frame. To simplify the presentation, 
	in the following we evaluate $q$ and $\etab$ in the LCMS.
	The acronym  LCMS stands for the Longitudinal Center of Mass System, 
	where the mean momentum of a 
	particle pair has vanishing longitudinal
	component, $K_z = 0.5 (k_{1,z} + k_{2,z}) = 0.$
	In this frame, 
	introduced in ref.~\cite{lcms}, 
	${\bf K}$ is orthogonal to the
	beam axis, and the time-like information on
	the duration of the particle emission couples
	to the out direction.
	The rapidity of the LCMS frame can be easily found
	from the measurement of the momentum vectors
	of the particles. As $\etab$ is from now on a space-time rapidity
	measured in the LCMS frame,
	it is invariant to longitudinal boosts:
	$\etab^\prime = (\etab - y) -  (0-y) = \etab$.

        The symbolic notation for the {\it side} direction is
	two dots  side by side as in $Q_{..}$.
	The remaining transverse direction, the {\it out} 
	direction was indexed as in $Q_{:}$,
	in an attempt to help to distinquish the zero-th component of the
        relative momentum $Q_0$ from the out component of the
        relative momentum $Q_{:} \equiv Q_o = Q_{out}$, $Q_0 \ne Q_o$.
	Hence $ K_{:} = |{\bf K}_\perp|$ and 
	$K_{..} = 0$.  The geometrical idea behind this notation is
        explained in details in ref.~\cite{3d-cf98}.
	The perpendicular (or transverse) component
	of the relative momentum is denoted by $Q_{\perp}$.
	By definition, $Q_{..}$, $Q_{:}$ and $Q_{\perp}$ are 
	invariants to longitudinal boosts, and $Q^2 = - q\cdot q =
	Q_{..}^2 + Q_:^2 + Q_{||}^2 - Q_=^2$. 

	A further simplification is obtained if we assume that
	the emission (or Wigner) function factorizes as  a product of 
	an {\it effective} proper-time distribution, a space-time rapidity
	distribution and a transverse coordinate distribution~\cite{lcms,3d}: 
\bea
	S_c(x,K) d^4 x &  = & H_*(\tau) G_*(\eta) I_*(r_x,r_y) \times 
			\nonumber \\
		&& \hspace{1.5truecm} \, d\tau \, \taub d\eta dr_x dr_y .\label{e:fact}
\eea
	The subscript $*$ stands for a dependence on the
	mean momentum $K$, the mid-rapidity $y_0$ and the 
	scale of violation of  boost-invariance $\Delta \eta$,
	using the symbolic notation $f_* \equiv f[K, y_0, \Delta\eta]$. 
	The function $H_*(\tau) $ stands for such an effective
	proper-time distribution (that includes, by definition,
	an extra factor $\tau$ from the Jacobian
	$d^4x = d\tau\, \tau\, d\eta, dr_x dr_y$,
	in order to relate
	the two-particle Bose-Einstein correlation function
	to a Fourier-transformation of a distribution function in $\tau$).
	The effective  space-time rapidity distribution
	is denoted by $G_*(\eta) $, while the  effective 
	transverse distribution is denoted by  $I_*(r_x,r_y) $ .
	In eq.~(\ref{e:fact}), the mean value of the proper-time $\taub$ is
	factored out, to keep the distribution functions dimensionless.
 
	With the help of the {\it small source size} 
	(or large relative momentum)
	expansion of eq.~(\ref{e:blexp}), 
	the amplitude 
	$\tilde s^i_c(1,2) \propto \langle a^{\dagger}_1 a_2 \rangle $
	that determines the arbitrary order Bose-Einstein correlation
	functions can be written  as follows: 
\be
	\tilde s^i_c(1,2) = \frac{ 
	\tilde H_*(Q_=) \tilde G_*(Q_{\parallel})  \tilde I_*(Q_{:},\, Q_{..})}
	{\tilde H_*(0) \tilde G_*(0) \tilde I_*(0,0) }. 
	\label{e:blamp}
\ee
	The Fourier-transformed distributions read as 
\bea
	\tilde H_*(Q_=)  & = & 
		\int d\tau e^{i Q_= \tau} H_*(\tau),
			\label{e:htild} \\
	\tilde G_*(Q_\parallel)  & = & \int
	d\eta 
		e^{- i Q_\parallel \taub \eta } G_*(\eta),
			\label{e:gtild}\\
	\tilde I_*(Q_:,Q_{..}) 
			 & = & \int dr_{:} dr_{..} 
			e^{- i Q_: r_: - i Q_{..} r_{..}} 
			I_*(r_:,r_{..}) . \nonumber  \\
	&& \label{e:itild}
\eea

	Utilizing the core-halo picture~\cite{cs-nato,chalo},
	the two-particle BECF 
	can be written into a factorized Buda-Lund form  as
\bea
	C({\bf k}_1, {\bf k}_2) & = & 1 + \nonumber \\
	&& \hspace{-2cm} \lambda_*
		{\dst |\tilde H_*(Q_=)  |^2 \ov |\tilde H_*(0) |^2} \,
		{\dst |\tilde G_*(Q_{\parallel})  |^2 \ov |\tilde G_*(0) |^2}\,
		{\dst |\tilde I_*(Q_{:},\, Q_{..})|^2 \ov 
		|\tilde I_*(0,0) |^2} .
	\label{e:blf}
\eea
	Thus, the BL results are rather generic. For example,
	BL parameterization may in particular limiting cases yield the  
	{\it power-law}, the {\it exponential}, the {\it double-Gaussian}, 
	the {\it Gaussian}, or the less familiar {\it oscillating}
	forms of ref.~\cite{3d-cf98}.
	The {\it Edgeworth}, the {\it Laguerre}  or other similarly constructed
	low-momentum expansions~\cite{lagu} 
	can be applied to any of the 
	factors of one variable in eq.~(\ref{e:blf})
	to characterize these unknown shapes in a 
	really model-independent manner, relying only on the
	convergence properties of expansions in terms
	of complete orthonormal sets of functions~\cite{lagu},
	as discussed below.

	In a Gaussian approximation and assuming that
	$R_{:} = R_{..} = R_{\perp}$, the  Buda-Lund form of BECF
	reads as follows:
\be
	{C_2({\bf k}_1,{\bf k}_2) = }
	{1}
	{+} \lambda_* \,
	{ e^{ - R^2_= Q^2_=
		     - R^2_{\parallel} Q^2_{\parallel}
	             - R_{\perp}^2 Q_{\perp}^2 } } ,
	\label{e:gaussblperp}
\ee
	where the 5 fit parameters are	$\lambda_*$, $R_=$,
	$R_\parallel$, $R_\perp$ and the value of $\etab$
	that enters the definitions of $Q_=$ and $Q_\parallel$
	in eqs.~(\ref{e:q=},\ref{e:qpar}).
	The fit parameter  $R_=$ reads as $R$-timelike, and this variable
	measures a width of the proper-time distribution $H_*$. 
	The fit parameter $R_{\parallel}$ reads as $R$-parallel, it
	measures an invariant
	length parallel to the direction of the expansion.
	The fit parameter $R_{\perp}$ reads as 
	$R$-perpedicular or $R$-perp. 
	For cylindrically symmetric sources, 
	$R_{\perp}$ measures a 
	transversal rms radius of the particle emitting source.

	The BL radius parameters characterize 
	the lengths of homogeneity~\cite{sinyukov} 
	in a longitudinally boost-invariant manner.
	The lengths of homogeneity are generally smaller 
	than the momentum-integrated, total extension of the source, 
	they measure a region in space and time, where particle pairs  
	with a given mean momentum  ${\bf  K}$ are emitted from. 

	In eq.~(\ref{e:gaussblperp}), 
	the spatial information about the source 
	distribution in $(r_x,r_y)$ was combined 
	to a single  perp radius parameter $R_{\perp}$. 
	In a more general Gaussian form, 
	suitable for studying rings of fire and 
	opacity effects, the \BL ~invariant
	BECF can be denoted as
\be
	C_2(q,K) = 1 + \lambda_* e^{ - R_=^2 Q_=^2 -
			 R^2_{\parallel}  Q^2_{\parallel} -
			R_{..}^2 Q_{..}^2 - R_:^2 Q_:^2 }.
		\label{e:bl-fring}
\ee
	The 6 fit parameters are $\lambda_*$, $R_=$,
	$R_{\parallel}$, $R_{..}$, $R_:$ and $\etab$,
	all are in principle functions of $({\bf K},y_0,\Delta\eta)$.
	Note, that this equation is identical to eq. (44) of
	ref.~\cite{3d},  rewritten into the new, symbolic notation
	of the Lorentz-invariant directional decomposition. 

	The above equation may be relevant for a study of
	expanding shells, or rings of fire, as discussed first in
	ref.~\cite{3d}.
	We argued, based on a simultaneous analysis of
	particle spectra and correlations, 
	and on recently found exact solutions of 
	non-relativistic fireball hydrodynamics~\cite{sol}
	that an expaning, spherical shell of
	fire is formed by the protons in 30 AMeV $^{40}Ar + ^{197}Au$
	reactions, and that a two-dimensional, expanding ring of fire
	is formed in the transverse plane in 
	NA22 $h+p$ reactions at CERN SPS~\cite{cs-nato}.  
	
	Opacity effects, as suggested recently 
	 by H. Heiselberg~\cite{henning}, 
	also require the distinction between $R_{..}$ and $R_{:}$. 
	The lack of transparency in the source may result in an 
	effective source function, that looks like a crescent
	in the side-out reference frame~\cite{henning}. 
	When integrated over the direction of the mean momentum,
	the effective source looks like a ring of fire in the
	$(r_x,r_y)$ frame. 

	The price of the invariant decomposition of the basic building
	blocks of any order Bose-Einstein correlation functions
	in the BL parameterization is that the correlation functions
	cannot be directly binned in the BL variables, 
	as these can determined after the parameter $\etab$
	is fitted to the data -- so
	the correlation function has to be binned first in 
	some directly measurable relative momentum components,
	e.g. the (side,out,long) relative momenta in the LCMS frame.
	After fitting $\etab$ in an arbitrary frame, 
	the BECF can be rebinned into the BL form.
	
	Other, more conventional parameterizations of the two-particle
	Bose-Einstein correlation functions are known as the
	Bertsch-Pratt (BP) and the Yano-Koonin-Podgoretskii (YKP)
	parameterizations.	
	These parameterizations exists only in Gaussian forms for
	the two-particle BECF, while BL forms exist for non-Gaussian
	Bose-Einstein correlations of arbitrary number of bosons.
	Other advantages and drawbacks of the BP and the YKP forms
	as compared to the Gaussian version of BL were discussed in 
	detail in refs.~\cite{3d-cf98,cs-nato}.

%-----------------------------------------------------------
%	3. Edgeworth
\section{Model-independent analysis of short-range correlations}

\def\bk{{\bf k}}
\def\vp{{\bf k}}
\def\vq{{\bf q}} 
\def\vk{{\bf k}}
\def\vK{{\bf K}}
\def\vx{{\bf x}}
\def\vy{{\bf y}}
\def\uk{{|{\bf k}|}}
\def\De{\Delta\eta}
\def\Des{\Delta\eta_*}
\def\Dk{\Delta k}
\def\Dt{\Delta\tau} 
\def\Dy{\Delta y}
\def\t0{\tau_0}
\def\tl{\tau_L}
\def\ch{\cosh}
\def\sh{\sinh}
\def\bea{\begin{eqnarray}}
\def\eea{\end{eqnarray}}
\def\I#1{\int d^3#1}
\def\bx{{\bf{x}}}
\def\bp{{\bf{p}}}
\def\bk{{\bf{k}}}
\def\bK{{\bf{K}}}
\def\bK{{\bf{K}}}
\def\bQ{{\bf{Q}}}
\def\bpi{{\bf{\pi}}}
\def\bxi{{\bf{\xi}}}
\def\bq{{\bf{q}}}
\def\br{{\bf{r}}}
\def\axd{\hat{ a}^{\dag} (\bx)}
\def\apd{\hat a^{\dag} (\bp)}
\def\ax{\hat{ a}^{} (\bx)}
\def\ap{\hat  a^{} (\bp)}
\def\ri{\right)}
\def\lef{\left(}
\def\dst{\displaystyle\phantom{|}}
\def\ov{\over\dst}
\def\om{\omega}
\def\eps{\epsilon}
\def\bak{{\bf K}}
\def\dek{{\bf \Delta k}}
\def\xb{{\overline{x}}}
\def\tb{{\overline{t}}}
\def\rb{{\overline{r}}}
\def\nb{{\overline{n}}}
\def\etab{{\overline{\eta}}}
\def\taub{{\overline{\tau}}}
%
% Now come the definitions for the new notation scheme
%
\def\D{\Delta}
\def\De{\Delta\eta}
\def\t{{\tau}}
\def\ch{\cosh}
\def\sh{\sinh}
\def\ben{\begin{eqnarray}}
\def\enn{\end{eqnarray}}
\def\bea{\begin{eqnarray}}
\def\eea{\end{eqnarray}}
\def\be{\begin{equation}}
\def\ee{\end{equation}}
\def\ov{\over\displaystyle\strut}
\def\dst{\displaystyle\phantom{|}}
\def\l({\left(}
\def\r){\right)}
\def\o{{out}}
\def\s{{side}}
\def\e{{\eta}}
\def\bdk{{\bf \Delta k}}
\def\rl{R_l^2}
\def\ro{R_{o}^2}
\def\rs{R_{s}^2}
\def\rol{R_{ol}^2}
\def\rpa{R_{\parallel}^2 }
\def\rpe{R_{\perp}^2 }
\def\rta{R_{\tau}^2 }
\def\BL{{Buda}{-}{Lund}~} 

The invariant Buda-Lund form corresponds to a small source size, 
``large" relative momentum expansion for the basic building block
of the Bose-Einstein correlation functions.
So it is natural to apply an expansion technique that is 
based on correction terms at small relative momentum, corresponding to
 large source sizes. This will be reviewed below following the lines
of refs.~\cite{lagu,cs-nato}.
   
The reviewed method is {\it really} model-independent, and it can be
applied not only to Bose-Einstein correlation functions
but to every experimentally determined function,
which features the properties {\it i)} and {\it ii)} listed below.

The following \underline{\it experimental properties} are assumed:

{\it i) } 
The measured function tends to a constant
for large values of the relative momentum.

{\it ii)}
The measured function has a non-trivial structure at
a certain value of its argument. 

The location of the non-trivial structure in the correlation
function is assumed for simplicity
to be close to $Q = 0$.

The properties {\it i)} and {\it ii)}
are well satisfied by e.g. the 
conventionally used two-particle Bose-Einstein
correlation functions.
For a critical review on the 
non-ideal features of short-range correlations,
(e.g. non-Gaussian shapes in multi-dimensional 
Bose-Einstein correlation studies),
we recommend ref.~\cite{kittel}.

The core/halo intercept parameter $\lambda_*$
is defined as the {\it extrapolated} value of the
two-particle correlation function at $Q = 0$, 
see ref~\cite{cs-nato} for greater details. It turns out, that
$\lambda_*$ is an important physical observable, 
related to the degree of partial restoration of $U_A(1)$ symmetry
in hot and dense hadronic matter~\cite{vck,ckv}.

A really model-independent 
 approach is to expand the measured
correlation functions in an abstract Hilbert space of functions.
It is reasonable to formulate such an expansion so that already
the first term in the series be as close to the
measured data points as possible. This can be achieved if one
identifies~\cite{lagu,edge-cst} the approximate shape
(e.g. the approximate Gaussian or the exponential shape) 
of the correlation function with the abstract measure
$\mu(t)dt$ in the abstract Hilbert-space ${\cal H}$. 
The orthogonality of the basis functions $\phi_n(t)$ in
${\cal H}$ can be utilized to guarantee the convergence
of these kind of expansions, see refs.~\cite{lagu,edge-cst}
for greater details.

\subsection{Laguerre expansion
\label{s:lagu}}
If in a zeroth order approximation the correlation function
has an exponential shape, then it is an efficient method to 
apply the Laguerre expansion, as a special case of
the general formulation of ref.~\cite{lagu,edge-cst}: 
\bea
C_2(Q)\! & =& \!{\cal N} \left\{ 
	1 + \lambda_L \exp(- Q R_L)  \right. \times \nonumber \\
	&& \hspace{-1truecm} \left.
	\left[ 1 + c_1 L_1(QR_L) + \frac{c_2}{2!} L_2(Q R_L) + ... \right]
	\right\}. 
	\nonumber \\
	&& \label{e:laguerre}
\eea
In this and the next subsection, $Q$ stands symbolically
for any, experimentally chosen, 
one dimensional relative momentum variable. 
The fit parameters are the scale parameters ${\cal N}$,
$\lambda_L$, $R_L$ and the expansion coefficients $c_1$, $c_2$, 
\, ... \, .  
The order $n$ Laguerre polynomials are defined as
\bea
	L_n(t) & = & \exp(t) \frac{d^n}{dt^n} t^n \exp(-t), 
\eea
they form a complete orthogonal basis for an exponential measure as
\bea
	\int_0^{\infty} dt \, \exp(-t) L_n(t) L_m(t)
	\propto
	\delta_{n,m} .
\eea
The first few Laguerre polynomials are explicitly given as
\bea
	L_0(t) & = & 1, \\
	L_1(t) & = & t  - 1,\\
	L_2(t) & = & t^2 - 4t + 2, \, ... \, .
\eea
As the Laguerre polynomials are non-vanishing at the origin,
$C(Q = 0) \ne 1 + \lambda_L$.
The physically significant core/halo intercept
parameter $\lambda_*$
can be obtained from the 
parameter $\lambda_L$ of the Laguerre expansion as
\bea
	\lambda_* & = & \lambda_L [1 - c_1 + c_2 - ... ] . 
\eea

\subsection{Edgeworth expansion 
\label{s:edge}}
If, in a zeroth-order approximation, the correlation function
has a Gaussian shape, then 
the general form given in ref.~\cite{edge} takes the particular form
of the Edgeworth expansion~\cite{edge-cst,edge,edge0} as:
\bea
C(Q) & = &{\cal N} \left\{ 
	1 + \lambda_E \exp( - Q^2 R_E^2) 
		\right. \times \nonumber \\
		&& 
	\left. 
	\hspace{-1.7truecm}
	\left[ 1 + \frac{\kappa_3}{3!} H_3(\sqrt{2} Q R_E)
		+\frac{\kappa_4}{4!} H_4(\sqrt{2} Q R_E)  ... \right]
		\right\} .
		\label{e:edge}
\eea
The fit parameters are the scale parameters ${\cal N}$,
$\lambda_E$, $R_E$ and the expansion coefficients $\kappa_3$, $\kappa_4$, 
\, ... \, ,  that coincide with the cumulants of rank 3, 4, ...,
of the correlation function.
The Hermite polynomials  are  defined as 
\bea
	H_n(t) & = & \exp( t^2/2) \left( - \frac{d}{dt} \right)^n
		\exp(-t^2/2), 
\eea
they form a complete orthogonal basis for an Gaussian measure as
\bea
	\int_{-\infty}^{\infty} dt 
	\, \exp(-t^2/2) H_n(t) H_m(t)
	 &  \propto  & \delta_{n,m}. 
\eea
The first few Hermite polynomials are listed as
\bea
	H_1(t) & = & t, \\
	H_2(t) & = & t^2 -1, \\
	H_3(t) & = & t^3 - 3 t , \\
	H_4(t) & = & t^4 - 6 t^2 + 3,\, ... \,
\eea
The physically significant core/halo intercept
parameter $\lambda_*$ can be obtained from the 
Edgeworth fit of eq.~(\ref{e:edge}) as
\bea
\lambda_* & = & \lambda_E \left[ 1 + \frac{\kappa_4}{8} + ... \right].
\eea
This expansion technique was applied  in the conference
contributions~\cite{edge-cst,edge} to the AFS minimum bias 
and 2-jet events to characterize
successfully the deviation of data from a Gaussian shape.
It was also successfully applied to characterize the 
non-Gaussian nature of the correlation function in two-dimensions
in case of the preliminary E802 data in ref.~\cite{edge-cst},
and it was recently applied to characterize the non-Gaussian nature
of the three-dimensional
two-pion BECF in $e^+ + e^-$ reactions at LEP1~\cite{l3-radii}.

	Fig.~1 of ref.~\cite{lagu}  indicates the ability of the
	Laguerre expansions to characterize two well-known,
	non-Gaussian correlation functions~\cite{lagu}: the 
	second-order short-range correlation function $D^s_2(Q)$ 
	as determined by the UA1 and 
	the NA22 experiments~\cite{ua1,na22-pwl}.  The fit results
	were summarized in Table 1 of ref.~\cite{lagu}.

From the fit results, the core/halo model intercept parameter is obtained as
$\lambda_* = 1.14 \pm 0.10$ (UA1) and $\lambda_* = 1.11 \pm 0.17$
(NA22). As both of these values are within errors equal to unity,
the maximum of the possible value of the intercept parameter
$\lambda_*$ in a fully chaotic source, we concluded~\cite{lagu} that
either there are other than Bose-Einstein short-range correlations
observed by both collaboration, or in case of this measurement
the full halo of long lived resonances is resolved~
\cite{chalo,dkiang,sk,bialas}.

If the two-particle BECF can be factorized as a product of (two or more)
functions of one variable each, then the Laguerre and the Edgeworth
expansions can be applied to the multiplicative factors -- functions of
one variable, each,
as done in refs.
~\cite{l3-radii,lagu,edge-cst,edge}.
The full, non-factorized form of  two-dimensional 
Edgeworth expansion and the interpretation
of its parameters is described in the handbook on mathematical statistics
by Kendall and Stuart~\cite{kendallstuart}.

%\vfill\eject

\section{Binary source formalism} 
	The first experimental evidence for
	oscillating behaviour in the two-particle correlation function
	has been observed by the NA49 collaboration, 
	unexpectedly, in the proton-proton correlations
	in central Pb+Pb collisions at CERN SPS  ~\cite{na49-pp}.
	The frequency of the oscillations has been explained
	with the help of the binary structure of the proton source
	in ref.~\cite{cs-nato}. Here we follow the lines of this
	presentation to explain the relationship between binary
	sources and oscillations in two-particle interferometry. 
	
	Binary sources appear
	generally: in astrophysics, in form of binary stars,
	in particle physics, in form of $W^+W^-$ pairs,
	that separate before they decay to hadrons.
 
	Let us consider first the simplest possible example, 
	to see how the binary sources result in
	oscillations in the Bose-Einstein or Fermi-Dirac
	correlation function. Suppose a 
	source distribution $s(x - x_+)$ describes, for example,
	a Gaussian source centered on $x_+$. 
	Consider a binary system, where the emission
	happens from $s_+= s(x-x_+)$ with fraction $f_+$,
	or from a displaced source, $s_-= s(x-x_-)$, 
	centered on $x_-$, with a fraction $f_-$. 
	For such a binary source, the amplitude of the emission is
	\be
		\rho(x) = f_+ s(x - x_+) + f_- s(x - x_-),
	\ee
	and the normalization requires
	\be
		f_+ + f_- = 1
	\ee
	The two-particle Bose-Einstein or Fermi-Dirac
	 correlation function is 
	\bea
		C(q) & = & 1 \pm | \tilde\rho(q)|^2 =
		1 \pm  \Omega(q) |\tilde s(q)|^2,
	\eea
	where $+$ is for bosons, and $-$ for fermions.
	The oscillating pre-factor $\Omega(q)$ satisfies
	$0 \le \Omega(q) \le 1$ and $\Omega(0) = 1$. This factor is
	given as 
	\be
	\Omega(q)  = 
		\left[ (f_+^2 + f_-^2) + 2 f_+ f_- \cos[q(x_+ - x_-)] \right]
	\label{e:omega}
	\ee
	The strength of the oscillations is controlled by
	the relative strength of emission from the displaced
	sources and the period of the oscillations
	can be used to learn about the distance of the 
	emitters. In the limit of one emitter ($f_+ = 1$ and
	$f_- = 0$, or  vice versa), the oscillations disappear.

	In particle physics, the effective separation between the
        sources can be estimated from the uncertainty
	relation to be  $x_\pm = |x_+ - x_-| \approx 
	2 \hbar/M_W \approx 0.005$ fm. Although this  is much
	smaller, than the effective size  of the pion source, 1 fm,
	one has to keep in mind that the back-to-back momenta
	of the $W^+ W^-$ pairs can be large, as compared to the pion mass.
	Due to this boost, pions with similar momentum may be emitted from
	different $W$-s with a separation 
	which is already comparable to the 1 fm hadronization scale,
	and the resulting oscillations may become observable.

        In stellar astronomy, the separation between 
	the binary stars is typically much larger than the
        diameter of the stars, hence the oscillations
	are well measurable. In principle, similar oscillations
	may provide a tool to measure the separation of the
	$W^+$ from $W^-$ in 4-jet events at LEP2.
	The scale of separation of $W^+W^-$ pairs
	is a key observable to estimate 
	in a quantum-mechanically correct manner the
	influence of the Bose-Einstein correlations
	on the reconstruction of the $W$ mass.	

\section{Oscillations in two-proton correlations }

	In heavy ion physics, oscillations are seen in the 
	short range of the $p + p$ correlation
	function~\cite{na49-pp}, with a half-period of $Q_h = 30$ MeV,
	in Pb+Pb reactions at CERN SPS measured by the NA49 collaboration.
	Proton-proton short-range correlations are strongly influenced 
	not only by the Fermi-Dirac statistics but also by Coulomb and
	strong final state interactions. This implies that the proton-proton
	correlation function of a binary proton source is approximately
	given by
\bea
	C^{pp}(q) & \simeq & 
	G(q) [1 + \Omega(q) ( C_s^{0,pp}(q) - 1)]
	\label{e:pposc}
\eea
	where $C^{0,pp}_s(q)$ stands for the two-proton
	correlation function including only the effects of
	strong final state interactions for one component of the
	binary source, $s(x)$. This  function $C_s^{0,pp}(q)$ 
	can be evaluated similarly to 
	the calculations performed for the conventional,
	one component, Gaussian like sources, see e.g.
	ref.~\cite{johann}, with the help of 
	the proton-proton relative wave-function 
	$\phi^{0}_{\bf q}({\bf x})$ which deviates
	from a relative plane wave due to the strong final state 
	interactions.  In eq.~(\ref{e:pposc}), 
	the binary nature of the source is represented by
	the oscillating pre-factor $\Omega(q)$
	the Columb final state interactions
	are represented in a small source size approximation
	by a conventional Gamow penetration factor $G(q)$.
	The above approximation assumes 
	that the strong final state interactions
	act on a scale that is smaller than the scale of separation of the
	binary proton source, hence the form of the oscillationg pre-factor
	$\Omega(q)$ is given by eq.~(\ref{e:omega}).
	The 30 MeV half-period of the oscillations in NA49 data
	implies a separation of $x_{\pm} = \pi \hbar/Q_h \approx 20$ fm,
	which is indeed much larger than the typical range of strong 
	final state interactions.  Hence the oscillations in $C_2^{pp}$
	can be attributed to interference between the
	the two peaks of the NA49 proton $dn/dy$ distribution
	~\cite{na49-p}, separated by $\Delta y = 2.5$.
	As for the protons we have $m >> T_0 = 140$ MeV,
	we can identify this rapidity difference with the
	space-time rapidity difference between the two
	peaks of the rapidity distribution. 
	The longitudinal scale of the separation
	is then given by $x_{\pm} = 2 \taub \sinh(\Delta\eta_p/2) $,
	which can be used to estimate the mean  freeze-out time
	of protons, $\taub =  \pi \hbar /[2 Q_h \sinh(\Delta\eta_p/2)] 
	\approx 6.4 $ fm/c, in a good agreement with the
	average value of $\taub = 5.9 \pm 0.6$ as extracted
	from the simultaneous analysis of the single-particle
	spectra and HBT radii in NA44, NA49 and WA98 experiments
	in the Buda-Lund picture, as summarized in
	refs.~\cite{ster-qm99,cs-nato}.	
	
	Note that the $^2He$ resonance is responsible
	for magnifying the structure of the two-proton
	correlation function at $Q \approx 50 - 100 $ MeV.
	This magnification makes the oscillations visible, 
	in spite of their relatively large half period.
	Another fortunate development in Pb+Pb collisions
	was that the $dn/dy$ distribution of the net baryon
	number (protons) developed two separate peaks, indicating
	perhaps the onset of nuclear transparency at 160 AGeV.
	At the smaller AGS bombarding energies, the 
	net baryon number has a single maximum at mid-rapidity
	which implies a very small, almost vanishing
	separation of the proton source  into two effective
	components, that correspond to unobservably large 
	period of oscillations, $\Omega \approx 1$ within the
	resolvable relative momentum range.

%-----------------------------------------------------------
%	8. Hydro models/ scaling behaviour
\section{
Oscillating Bose-Einstein correlations}
\label{s:bl-h}

	The two-particle Bose-Einstein correlation function of the
	Buda-Lund hydro model  (BL-H)
	was evaluated in ref.\cite{3d} in a Gaussian approximation,
	without applying the binary source formulation.
	An improved calculation was 
 	presented in ref.~\cite{3d-cf98}, where the
	correlation function was evaluated 
	using in the binary source formulation,  
	and the corresponding oscillations were found.

	Using the exponential form of the $\cosh[\eta -y]$ factor,
        the BL-H emission function $S_c(x,{\bf k}) $
	can be written as a sum of two terms:
\bea
        S_c(x,{\bf k}) & = & 0.5 [ S_+(x,{\bf k}) + S_-(x,{\bf k}) ] ,\\
        S_\pm(x,{\bf k}) & = & \frac{g}{(2 \pi)^3}
	\frac{ m_t \exp[\pm\eta \mp y]  H_*(\tau) }
	{[f_B(x,{\bf k}) + s]} 
		,\\
        f_B(x,{\bf k}) & = &\exp\left[\frac{k^{\mu}  u_{\mu}(x) 
				- \mu(x)}{T(x)} \right] ,
\eea
	and $s = 0, \pm 1$ for Boltzmann, Fermi-Dirac or Bose-Einstein
	distributions.

	The above splitting is the basis of the binary source formulation
	of the BL-H parameterization.
        The effective emission function components are both subject to
        Fourier - transformation in the BL approach.
        In an improved saddle-point approximation, the two components
        $S_+(x,k)$ and $S_-(x,k)$ can be Fourier - transformed
        independently, finding the separate
        maxima (saddle point) $\xb_+$ and $\xb_-$ of
        $S_+(x,k)$ and $S_-(x,k)$, and performing the analytic 
        calculation for the two components separately.

        These oscillations in the intensity correlation function are 
        similar to the oscillations in the intensity correlations
        of photons from binary stars in stellar astronomy~\cite{hbt-bin}.

        Note that the oscillations are expected to be small in the BL-H
	picture, and the
        Gaussian remains a good approximation to the BECF
        but with modified radius parameters.

	The Buda-Lund hydro parameterization (BL-H)
	was invented in the same paper as the
	BL parameterization of the
	Bose-Einstein correlation functions~\cite{3d},
	but in principle the general BL forms of the 
	correlation function do not depend on the hydrodynamical
	ansatz (BL-H).

	The BL-H  two-particle Bose-Einstein correlation function
	is evaluated in the binary source formalism in 
	ref.~\cite{3d-cf98}:
\be
        C_2({\bf k}_1,{\bf k}_2) =
	%\!\!\! & = &\!\!\!
        1 +
         \lambda_* \, \Omega \,\,\,
                e^{- Q_{\parallel}^2 \overline{R}_{\parallel}^2
                 - Q_{=}^2 \overline{R}_{=}^2
                 - Q_{\perp}^2 \overline{R}_{\perp}^2 }, 
	\nonumber
\ee
	where the pre-factor $\Omega$
	induces oscillations within the Gaussian envelope as
	a function of $Q_\parallel$. This  oscillating pre-factor satisfies
	$0 \le \Omega(Q_\parallel) \le 1$ and $\Omega(0) = 1$. This factor is
	given as
\bea
        \Omega  & = &   \cos^2( Q_{\parallel} 
			\overline{R}_{\parallel} \,
       { \Delta\etab} ) \times  \nonumber \\
	&& 
	\left[ 1 + \tan^2(Q_{\parallel} \overline{R}_{\parallel} \,
			{ \Delta\etab})
                  \tanh^2\etab \right] . 
\eea
	The invariant BL decomposition of the relative 
	momentum is utilized  to present the correlation function
	in the simplest possible form.  
        Although the shape of the BECF is non-Gaussian,  
	because the factor $\Omega(Q_{\parallel})$ 
	results in oscillations of the correlator,
        the result is still explicitely boost-invariant.
	Although the source is assumed to be cylindrically
	symmetric, we have 6 free fit parameters
	in this BL form of the correlation function:
	$\lambda_*$, $R_=$, $R_{\parallel}$, $R_\perp$,
	$\etab$ and ${\Delta\etab}$.
	The latter two control the period of the oscillations
	in the correlation function, which in turn
	carries information on the separation of the
	effective binary sources. This emphasizes the
	importance of the oscillating factor in the 
	BL Bose-Einstein correlation function. 

        Due to the analytically found oscillations,
	the presented form of the BECF
	goes beyond the single Gaussian version of the saddle-point
        calculations of ref.~\cite{uli_s,uli_l}.
        This result goes also beyond the results obtainable in
        the YKP  or the BP parameterizations. In principle,
        the binary-source saddle-point calculation gives more accurate
        analytic results than the numerical evaluation of space-time variances,
        as the binary-source calculation
	keeps non-Gaussian information on the detailed shape 
	of the Bose-Einstein correlation function.
	The parameters of the spectrum and the correlation 
	function are the same,  given in more details in
	ref.~\cite{cs-nato}. Hence the simultaneous analysis of the
	two-particle correlations and the single-particle spectra,
	advocated already in refs.~\cite{3d}, yields a 
	rather precise picture of particle  production.
	Without the oscillations, it is possible to
	determine only the means and the variances of the
	density,  flow and temperature profiles of the particle source.
	By taking into account  the oscillations, additional information	
	about the separation of the source into an effective binary structure
	can also be established.

%-----------------------------------------------------------
%	16. Summary
\section{Summary and outlook}
  	One of the new directions in two-particle quantum 
	statistical correlation
	studies is to search for non-Gaussian structures .
	An important observation is that particle interferometry for binary 
	sources predicts the existence of oscillations in the two-particle
	Bose-Einstein or Fermi-Dirac correlation functions. 
	Intensity correlations oscillate, if the sources are binary,
	and the separation of the binary sources is related to the 
	period of the oscillations. Novel model-independent tools to
	search for such oscillatory patters are the Edgeworth or the Laguerre
	expansion techniques. 
	The frequency of
	oscillations in the two-proton correlation function in Pb + Pb
	collisions at CERN SPS has been explained in terms of the BL-H model
	as a consequence of the separation of the proton sources 
	in this reaction.

	More work is required to work out the greater details of oscillatiory
	patterns in two-particle Bose-Einstein correlations.

\section*{Acknowledgments} 
	I would like to thank the Organizers of Torino 2000 
	for creating a pleasent athmosphere and an
	inspiring working  environment. 
	
	This research was supported by a Bolyai Fellowship
	of the Hungarian Academy of Sciences and by the grants OTKA
	T024094, T026435, T029158, the US-Hungarian Joint 
	Fund MAKA grant 652/1998, NWO - OTKA 
	N025186, OMFB - Ukraine S\& T grant 45014
	and  FAPESP 98/2249-4 and 99/09113-3. 

%\vfill\eject

\end{document}